%% file: ms.tex
\documentclass[final,3p]{elsarticle}

\usepackage[utf8]{inputenc}
\usepackage{hyperref}

\usepackage{amsmath,amsfonts,amsthm}
\usepackage{mathtools,commath,nicefrac}
\usepackage{booktabs,multirow}
\usepackage{etoolbox}
\usepackage{bm}
\DeclareMathAlphabet{\pazocal}{OMS}{zplm}{m}{n}
\usepackage{nicefrac}

\usepackage[textsize=scriptsize]{todonotes}
\usepackage{graphicx}
\usepackage{tikz}	
\usepackage{pgfplots}
	\pgfplotsset{compat=newest}
\usetikzlibrary{pgfplots.groupplots}

\pgfplotsset{every axis/.append style={
                    label style={font=\scriptsize},
                    axis line style = {line width=1pt}, 
                    tick label style={font=\scriptsize},
                    line width=0.7pt,
                    mark options={solid},
                    anchor=center,align=center
                    }
}

\newcommand{\OM}{\Omega}							
\newcommand{\FL}{\pazocal{F}}					
\newcommand{\SO}{\pazocal{S}}					
\newcommand{\IN}{\pazocal{I}}					

\newcommand{\Gin}{\Gamma_{\text{in}}}			
\newcommand{\Gout}{\Gamma_{\text{out}}}			
\newcommand{\Gwall}{\Gamma_{\text{wall}}}		

\newcommand{\R}{\mathbb{R}}						
\newcommand{\ub}{\mathbf{u}} 					
\newcommand{\uin}{\ub_\text{in}}					
\newcommand{\sigb}{\bm{\sigma}}					
\newcommand{\fb}{\mathbf{f}}						
\newcommand{\Fb}{\mathbf{F}}						

\newcommand{\vb}{\mathbf{v}}						
\newcommand{\Tb}{\mathbf{T}}						
\newcommand{\angV}{\bm{\omega}}					
\newcommand{\cs}{\mathbf{c}_\SO}					

\newcommand{\xb}{\mathbf{x}}						
\newcommand{\tang}{\vec{\mathbf{t}}}				
\newcommand{\norml}{\vec{\mathbf{n}}}			

\renewcommand{\div}{\operatorname{div}}			
\newcommand{\id}{\operatorname{Id}}				

\newcommand{\restr}[2]{{\left.\kern-\nulldelimiterspace #1
\right|_{#2} }}
\newcommand{\Poly}{\mathbb{P}}					

\newcommand{\icol}[1]{
  \left(\begin{smallmatrix}#1\end{smallmatrix}\right)%
}

\makeatletter
\renewcommand{\fps@figure}{htp}
\renewcommand{\fps@table}{htp}
\makeatother

\newtheorem*{remark}{Remark}

\journal{`Computers \& Fluids' (accepted: September 5, 2019)}

\begin{document}

\begin{frontmatter}

\title{Numerical benchmarking of fluid-rigid body interactions}

\author[ovgu]{Henry von Wahl\corref{cor1}}
\cortext[cor1]{Corresponding author}
\ead{henry.vonwahl@ovgu.de}
\fntext[fn1]{ORCID: \url{https://orcid.org/0000-0002-0793-1647}}

\author[ovgu]{Thomas Richter}
\fntext[fn2]{ORCID: \url{https://orcid.org/0000-0003-0206-3606}}

\author[goe]{Christoph Lehrenfeld}
\fntext[fn3]{ORCID: \url{https://orcid.org/0000-0003-0170-8468}}

\author[mpi,ovgu]{Jan Heiland}
\fntext[fn4]{ORCID: \url{https://orcid.org/0000-0003-0228-8522}}

\author[ovgu]{Piotr Minakowski}%
\fntext[fn5]{ORCID: \url{https://orcid.org/0000-0001-5154-2967}}

\address[ovgu]{Institute for Analysis and Numerics, Otto-von-Guericke-University, Universit\"atsplatz 2, 39106 Magdeburg, Germany}

\address[goe]{Institute for Numerical and Applied Mathematics, Georg-August-University, Lotzestr. 16-18, 37083 G\"ottingen, Germany}

\address[mpi]{Max Planck Institute for Dynamics of Complex Technical Systems, Sandtorstra\ss e, 39106 Magdeburg, Germany}

\begin{abstract}
	We propose a fluid-rigid body interaction benchmark problem, consisting of a solid spherical obstacle in a Newtonian fluid, whose centre of mass is fixed but is free to rotate. A number of different problems are defined for both two and three spatial dimensions. The geometry is chosen specifically, such that the fluid-solid partition does not change over time and classical fluid solvers are able to solve the fluid-structure interaction problem. We summarise the different approaches used to handle the fluid-solid coupling and numerical methods used to solve the arising problems. The results obtained by the described methods are presented and we give reference intervals for the relevant quantities of interest.
\end{abstract}


\noindent Publisher's version: DOI \href{https://doi.org/10.1016/j.compfluid.2019.104290}  {\texttt{10.1016/j.compfluid.2019.104290}}\\
\copyright~2019. This manuscript version is made available under the CC-BY-NC-ND 4.0 license: \\
\url{https://creativecommons.org/licenses/by-nc-nd/4.0/}

\end{frontmatter}


\input{intro}

\input{config}

\input{methods}

\input{results}

\input{back_matter}

\def\bibsection{\section*{References}}
\bibliography{bibexport}
\end{document}

%% file: intro.tex
\section{Introduction}

The interaction between a fluid flow and rigid bodies appears in many physical applications. The flow around a free rigid body causes both displacement and rotation of that body, via the forces and torque exerted from the fluid onto the body. Conversely, the motion of the body causes changes in the flow. In this work we focus on pure rotational effects of a single rigid sphere in two and three spatial dimensions by fixing its centre of mass. 

We consider the fluid-structure interaction between an incompressible Newtonian fluid and a rigid body, which is free to rotate around it's centre of mass. The geometric set-up is based on the well known computational fluid dynamics (CFD) benchmark \emph{Flow around a cylinder} \cite{SchaeferTurek1996}, while adding the coupling between a freely rotating body and the fluid. The main advantage of this set-up is the constant fluid-solid partition over time, which allows the use of well established CFD methods to compute this fluid-structure interaction (FSI) problem, placing the focus on the coupling/decoupling approach used between the solid and the fluid.

Well posed benchmark problems are vital within computational mathematics. They are important for code validation and comparison of methods. For example the benchmark problem \emph{Flow around a cylinder} from Sch\"afer and Turek \cite{SchaeferTurek1996} is one of the most widely used benchmarks within the computational fluid dynamics community. More recently Turek and Hron \cite{TurekHron2006} proposed an FSI benchmark between an elastic obstacle and a laminar flow, while Hysing et al \cite{HysingTurekKuzminParoliniBurmanGanesanTobiska2009} proposed a benchmark for two dimensional bubbles rising in a fluid. In this work, we propose a benchmark problem in the setting of fluid-rigid body interactions.

The main contribution of this paper is to present a set of benchmark problems and relevant benchmark quantities for a fluid-rigid body interaction problem which can be used to quantitatively establish the quality of methods for fluid-particle interactions. This benchmark is more relevant in this situation than standard CFD benchmarks, since we incorporate the coupling between the solid and the fluid while keeping the set-up simple to implement.

Rotating spheres and cylinders in a channel flow have been studied widely in the literature, e.g. Badr et al \cite{BadrDennisYoung89}, Fabre et al \cite{FabreTchoufagCitroGiannettiLuchini16}, Housiadas and Tanner \cite{HousiadasTanner11}, Juarez et al \cite{JuarezScottMetcalfeBagheri00}, Kong et al\cite{KangChoiLee99}, Mittal and Kumar \cite{MittalKumar03}, Shaafin et al \cite{ShaafiNaikVengadesan17}, Stojkovic et al \cite{StojkovicBreuerDurst02} and Xia et al \cite{XiaLinKuChan18}. The main focus has been the effect of different configurations and parameter choices on the resulting flow. Computations have usually been done using a single method, thus reliable reference results for a given set-up are not available. The flow past a cylinder in two dimensions, rotating at a given prescribed rate, has been studied (see for example \cite{KangChoiLee99,MittalKumar03,StojkovicBreuerDurst02}) with the emphasis on the appearance of vortex shedding at different rotation rates, dependent on the  Reynolds number. Furthermore, the effects of rotation speed and eccentricity (distance between the wall and cylinder) of the cylinder in two dimensions was studied \cite{ShaafiNaikVengadesan17} and the dynamics of a rotating cylinder moving at a constant speed \cite{BadrDennisYoung89}. The case of a two dimensional cylinder which is free to rotate was studied by Juarez et al\cite{JuarezScottMetcalfeBagheri00}, looking at the effects of eccentricity and Reynolds number on the rate and direction of rotation of the cylinder, while Xia et al \cite{XiaLinKuChan18} also took the blockage ratio into account. Three dimensional situations have also been considered in the literature \cite{FabreTchoufagCitroGiannettiLuchini16,HousiadasTanner11}. In Fabre et al \cite{FabreTchoufagCitroGiannettiLuchini16} a sphere rotating freely around a transverse axis in a Newtonian fluid was considered and numerical results were compared to the case of a fixed sphere, while Housiadas and Tanner \cite{HousiadasTanner11} studied a freely rotating sphere in a viscoelastic fluid and presented analytical results for the angular velocity.

The layout of this paper is as follows: In Section \ref{sec:Config} we begin by describing the governing equations of this benchmark. We then describe the two dimensional set-up in Section \ref{sec:Config-subsec:2Dsetup} and the three dimensional set-up in Section \ref{sec:Config-subsec:3Dsetup}, as well as defining the reference quantities which should be computed in each test case. In Section \ref{sec:NumMet}, we describe the different numerical methods and coupling/decoupling methods used to compute the described problems. The computed quantities of interest for the different benchmark set-ups and different methods are presented in Section \ref{sec:Results}.

%% file: config.tex
\section{Configurations of the benchmark problem}\label{sec:Config}

\subsection{Governing equations}\label{sec:Cinfig-subsec:Eqns}

Consider a finite time-interval $I=[0,t_{\text{end}}]$ and a bounded domain $\OM\in\R^d$ for $d\in\{2,3\}$. We split the domain into a $d$-dimensional fluid region $\FL$, a $d$-dimensional solid region $\SO$ and a $(d-1)$-dimensional interface $\IN$, such that $\OM=\FL\cup\SO\cup\IN$. In $\FL$ we prescribe the incompressible Navier-Stokes equations
\begin{equation}\label{eqn:NavierStokes}
	\begin{aligned}
		\rho_f\left(\partial_t\ub + (\ub\cdot\nabla)\ub\right) - \div\sigb(\ub,p) &= \rho_f\fb_f\\
		\div(\ub) &= 0
	\end{aligned}
\end{equation}
where $\ub:\FL\rightarrow\R^d$ denotes the fluid velocity field (note that we will denote vector valued functions with bold symbols), $p:\FL\rightarrow\R$ the pressure, $\rho_f$ the fluid's density, $\fb:\Omega\rightarrow\R^d$ an external forcing vector acting on the fluid and
\begin{equation*}
	\sigb(\ub,p) = \rho_f\nu\left( \nabla\ub+\nabla\ub^T \right) - p\id
\end{equation*} 
denotes the Cauchy stress-tensor, with the kinematic viscosity $\nu>0$. The fluid boundary is split into an inflow boundary $\Gin$, an outflow boundary $\Gout$, rigid no-slip wall boundaries $\Gwall$ and the fluid-solid interface $\IN$. On the inflow, wall and interface boundaries we impose Dirichlet boundary conditions while on the outflow condition we apply the do-nothing condition (see e.g. \cite{HeywoodRannacherTurek1992})
\begin{equation}\label{eqn:FluidBoundaryCond}
	\begin{aligned}
		\ub &= \uin 	&\text{on }& \Gin\times I\\
		\ub &= 0 		&\text{on }& \Gwall\times I\\
		\ub &= \vb_s 	&\text{on }& \IN\times I\\
		 \rho_f\nu \left(\norml\cdot\nabla\right)\ub- p\norml &=0 &\text{on }& \Gout\times I\\
	\end{aligned}
\end{equation}
where $\uin$ is a prescribed inflow-profile, $\vb_s$ is the solid's velocity at the fluid-solid interface and $\norml$ is the outward pointing unit normal vector. 

We assume that the rigid body $\SO$ has uniform density and the motion of $\SO$ is restricted to free rotation around its centre of mass. The motion can therefore be described by the object angular velocity $\angV$ (in a counter-clockwise sense), which is governed by Newton's second law of motion. In our case, this can be stated as an ordinary differential equation for the angular velocity, given by
\begin{equation}\label{eqn:RotationODE}
	J\partial_t\angV = \Tb
\end{equation}
where $J$ is the body's moment of inertia and $\Tb$ is the total torque exerted onto the body by the fluid. Note that in two dimensions \eqref{eqn:RotationODE} is a scalar equation. The total torque $\Tb$ is then given by
\begin{equation}\label{eqn:DefTorque}
	\Tb = \int_\IN (\xb-\cs)\times \left( \sigb( \ub,p )\norml \right)\dif s
\end{equation}
with the body's centre of mass $\cs$.
\begin{remark}
	Equations \eqref{eqn:RotationODE} and \eqref{eqn:DefTorque} are valid in three dimensions. In order to extend this formulation into two dimensions, we embed the the two dimensional domain trivially into three dimensions and take the scalar torque as the non-trivial component of the cross product in \eqref{eqn:RotationODE}. A simpler formulation for this will be given below. Similarly, the angular velocity is then also taken as the one non-trivial component of the three dimensional angular velocity.
\end{remark}

\subsection{Two-dimensional benchmark set-up}\label{sec:Config-subsec:2Dsetup}

The set-up is based on the 2d-benchmark problems defined by Schäfer and Turek\cite{SchaeferTurek1996} and is chosen such that the partition of $\OM=\FL\cup\SO\cup\IN$ does not change over time in a geometric sense. This allows the use of standard CFD codes to compute these FSI problems.

We choose $\OM=[0,2.2]\times[0,0.41]$ while the solid $\SO$ is a circular obstacle placed just below the vertical center of the domain $\SO = \{\icol{x\\y}\in\OM : \big\Vert \icol{x-0.2\\y-0.2}\big\Vert_2< R  \}$ with $R=0.05$. The inflow data is defined by a parabolic inflow profile
\begin{equation*}
  \uin(t) \coloneqq \frac{4U(t)y(0.41-y)}{0.41^2}\begin{pmatrix}1\\0   \end{pmatrix}
\end{equation*}
for some inflow speed $U$. The velocity at the interface is given by $\vb_s=\omega R\tang$, where $\tang$ is the unit tangential vector (pointing in the anti-clockwise direction). The fluid viscosity is set to $\nu=0.001$, the fluid density $\rho_f=1$ and the solid density $\rho_s=10$. In the case of a circle, the moment of inertia is then given by
\begin{equation*}
	J = \rho_s\int_{\SO}\vert\xb-\cs\vert^2\dif\xb = \rho_s\int_0^{2\pi}\int_0^Rr^2 r\dif r\dif\theta = \rho_s\frac{\pi}{2}R^4.
\end{equation*}
In the two dimensional case, \eqref{eqn:DefTorque} reduces to the scalar equation
\begin{equation*}
	T=\int_{\IN}R\tang\cdot\big(\sigb(\ub,p)\norml \big)\,\dif s
\end{equation*}
and as a result, the pressure does not contribute to changes in angular velocity.

As benchmark quantities, we are interested in the forces acting on the solid $\SO$. The forces acting on the solid in the horizontal and vertical directions are
\begin{equation}\label{eqn:DirectionalForces}
	\Fb = \int_\IN \sigb(\ub,p)\norml\dif s.
\end{equation}
As reference values, we then take the dimensionless drag (horizontal force, $x$-direction) and lift (vertical force, $y$-direction) coefficients which are defined as 
\begin{equation*}
C_D=\frac{2}{U^2_m\rho_fL}\Fb_1\quad\text{and}\quad C_L=\frac{2}{U^2_m\rho_f L}\Fb_2
\end{equation*}
with the mean inflow speed $U_m$ and characteristic length $L$, which we take to be the diameter of the circle, i.e., $L=0.1$. To characterise the rotational force we take the dimensionless torque coefficient
\begin{equation*}
	C_T = \frac{4}{U^2_m\rho_fL^2}T.
\end{equation*}
Furthermore, we compute the dimensionless angular velocity
\begin{equation*}
\omega^\ast = \frac{\omega L}{2U_m}
\end{equation*}
and the pressure difference between the front and the back of the solid
\begin{equation*}
	\Delta p  = p((0.15,0.2))- p((0.25,0.2)).
\end{equation*}
To characterise the fluid we take the Strouhal number, defined as $St=\nicefrac{Lf}{U_m}$,  with the frequency of vortex shedding $f$, and the Reynolds number which we take as $Re=\nicefrac{U_m L}{\nu}$.

\begin{figure}[t]
  \begin{center}
    \includegraphics[width=0.8\textwidth]{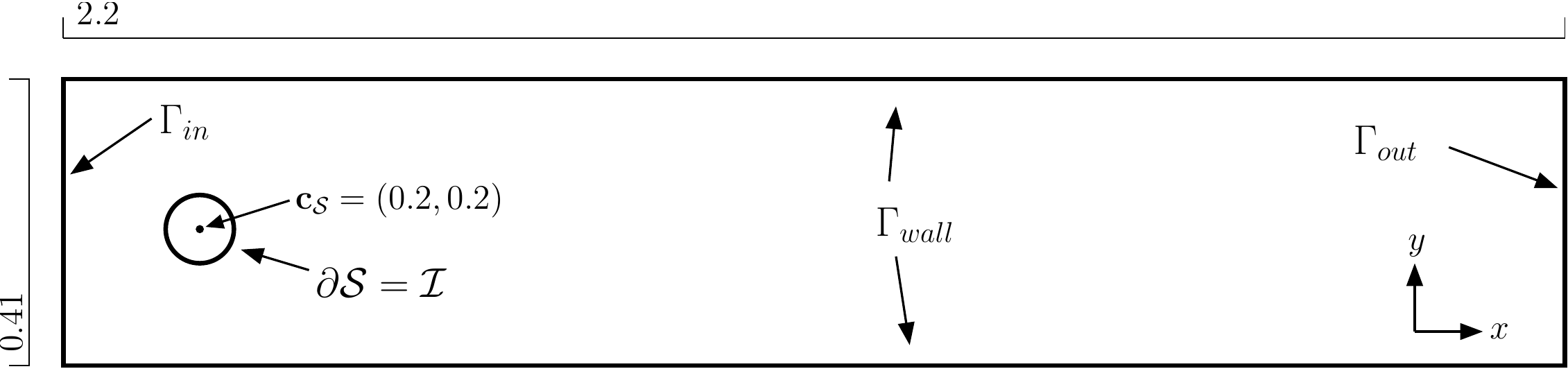}
    \caption{Spatial configuration of the two-dimensional benchmark problems.}
  \end{center}
\end{figure}

\subsubsection{Two-dimensional problem description}

As in Schäfer and Turek\cite{SchaeferTurek1996}, we consider three different two-dimensional test cases.

\paragraph{Rot2d-1 (Stationary)}

The inflow speed is set to $U=0.3$. This results in a mean inflow speed of $U_m=\nicefrac{2\cdot0.3}{3}=0.2$ which in turn gives the Reynolds number $Re=20$. 

The quantities to be computed are the drag, lift and torque coefficients $C_D$, $C_L$ and $C_T$, the pressure difference $\Delta p$ and the dimensionless angular velocity $\omega^\ast$. The reference values are the stationary limit values of these quantities.

\paragraph{Rot2d-2 (Periodic)}

The inflow speed is $U=1.5$, which gives a mean inflow speed of $U_m=1$ and the Reynolds number $Re=100$. 

To be computed are the drag, lift and torque coefficients $C_D$, $C_L$ and $C_T$, the pressure difference $\Delta p$ and the angular velocity $\omega^\ast$ over one period $[t_0,t_0+\nicefrac{1}{f}]$ where $f=f(C_D)$ is the frequency of the drag coefficient. The reference values here are the maximum and minimum drag, lift and torque coefficients $C_{D,\text{max}}$, $C_{D,\text{min}}$, $C_{L,\text{max}}$, $C_{L,\text{min}}$, $C_{T,\text{max}}$ and $C_{T,\text{min}}$, the maximum and minimum dimensionless angular velocity $\omega^\ast_\text{max}$ and $\omega^\ast_\text{min}$, the Strouhal number and the pressure difference $\Delta p(t^\ast)$ at $t^\ast=t_0+\nicefrac{1}{2f}$, the midpoint of one period. The "initial" time $t_0$ corresponds to the time at which the maximum of the lift coefficient is realised.

\paragraph{Rot2d-3 (Unsteady \& fixed time)}

We consider the fixed time interval $[0,8]$ and the inflow speed $U(t)=1.5\sin(\pi\nicefrac{t}{8})$. At the maximum in time, the resulting mean velocity (in space) is $U_m=1$ and Reynolds number is again $Re=100$. The initial state is $\ub(0) = \bm{0}$.

To be computed are the drag, lift and torque coefficients $C_D$, $C_L$ and $C_{T}$, the pressure difference $\Delta p$ and the angular velocity $\omega^\ast$. The reference values are the maximum drag, lift and torque coefficients $C_{D,\text{max}}$, $C_{L,\text{max}}$ and $C_{T,\text{max}}$ and the respective times at which these are realised, the maximum dimensionless angular velocity $\omega^\ast_\text{max}$ and the time at which it is realised as well as the pressure difference at time $t=8$.

\subsection{3D Set-up}\label{sec:Config-subsec:3Dsetup}

We consider the channel domain $\OM=[0,2.5]\times[0,0.41]\times[0,0.41]$ with a spherical obstacle $\SO = \{ \xb\in\OM : \Vert \xb-\cs \Vert_2<R\} $ with $\cs=(0.5,0.2,0.18)^T$ and $R=0.05$, which is again free to rotate around it's centre of mass, $\cs$. This geometry has again been chosen such that the fluid-solid partition does not change over time and standard CFD codes can be used to compute this FSI problem. The inflow data for this problem is given by
\begin{equation*}
  \uin(t) = \frac{16U(t)y(y-0.41)z(z-0.41)}{0.41^4}\begin{pmatrix}1\\0\\0
  \end{pmatrix}
\end{equation*} 
with a maximum inflow speed $U(t)$. In the three dimensional case, the angular velocity $\angV$ has three components, each of which represents the angular velocity around the corresponding axis. The coupling of the fluid and solid is then given by $\ub=\vb_s=\angV\times(\xb-\cs)$ on $\IN$. The fluid viscosity is set to $\nu=0.001$, the fluid density is taken as $\rho_f=1$ and for the solid we choose $\rho_s=10$. In the case of a sphere, the moment of inertia is
\begin{equation*}
	J= \frac{2}{5}mR^2
\end{equation*}
with the mass of the sphere being $m=\rho_s \nicefrac{4}{3}\,\pi R^3$. The torque is computed as in \eqref{eqn:DefTorque}.

As reference quantities for this benchmark, we again take the forces acting on the obstacle. The forces acting in the coordinate axis are again defined by \eqref{eqn:DirectionalForces}. As reference values we take the dimensionless coefficients, defined as
\begin{equation*}
	C_{\Fb_i} = \frac{2}{U^2_m\rho_fL^2}\Fb_i \quad\text{and}\quad C_{\Tb_i}= \frac{4}{U^2_m\rho_fL^3}\Tb_i
\end{equation*}
for $i\in\{1,2,3\}$, where we take the sphere diameter $L=0.1$ as our reference length. The dimensionless angular velocity around each coordinate axis is given by $\angV_i^\ast=\nicefrac{\angV_i L}{2 U_m}$. The pressure difference is defined as
\begin{equation*}
	\Delta p  = p((0.45,0.2,0.18))- p((0.55,0.2,0.18))
\end{equation*}
while the Strouhal and Reynolds numbers are defined as in the two-dimensional case.

\begin{figure}
  \begin{center}
    \includegraphics[width=0.8\textwidth]{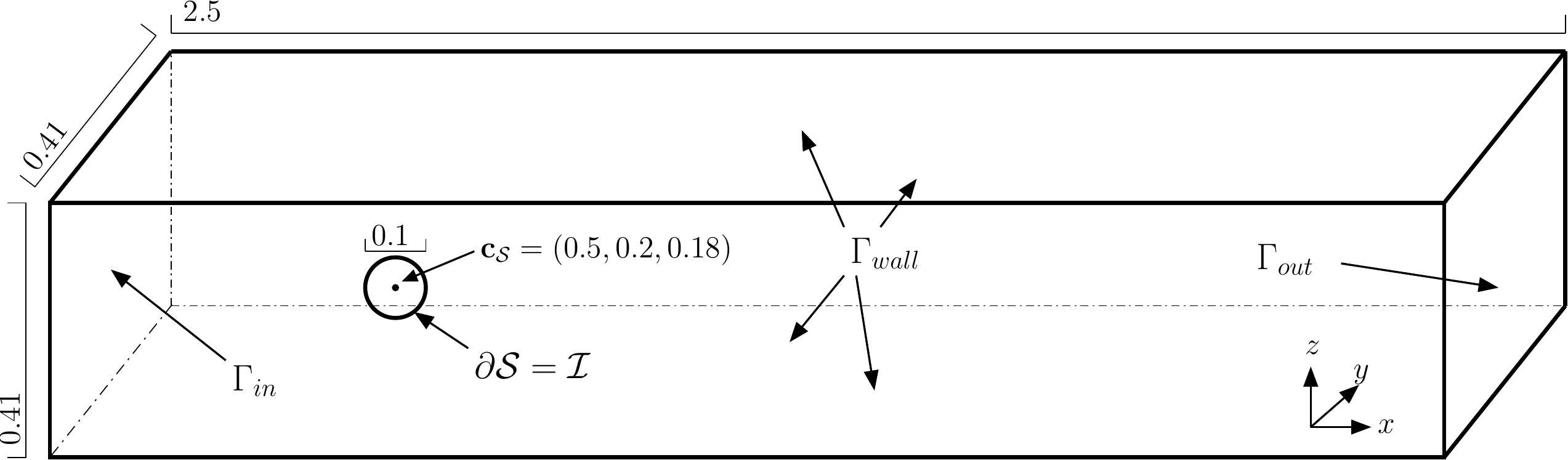}
    \caption{Spatial configuration of the three-dimensional benchmark problems. }
  \end{center}
\end{figure}

\subsubsection{Three-dimensional problem description}

We only consider one stationary case, as we were unable to find a suitable stable configuration at higher Reynolds numbers.

\paragraph{Rot3d-1 (Stationary)}

We choose $U=0.45$ giving a mean speed of $U_m=\nicefrac{4\cdot0.45}{9}=0.2$. As a result of this choice we get the Reynolds number $Re=20$.

To be computed are the three force coefficients $C_{\Fb_i}$, the Euclidean norm of the torque coefficient vector $\Vert C_{\Tb}\Vert_2$, the pressure difference $\Delta p$ and the three components of the dimensionless angular velocity $\angV^\ast$. The reference values are the stationary limits of these quantities.

%% file: methods.tex
\section{Numerical methods}\label{sec:NumMet}

We describe the different methods utilised here to compute the described problems.

\subsection{\texorpdfstring{Grad-div stabilised high-order Taylor-Hood (TH\textsubscript{gd,ho})}{
	Grad-div stabilised high-order Taylor-Hood}}\label{Sec:HvW}

We consider the conforming, inf-sup stable family of Taylor-Hood finite element pairs $\Poly^k/\Poly^{k-1}$ for $k\geq2$ \cite{Taylor1973,Boffi1994} together with grad-div stabilisation \cite{FrancaHughes1988}, i.e. we add the term
\begin{equation*}
	\gamma_{gd}(\div(\ub_h),\div(\vb_h))_{\OM}
\end{equation*}
to the (Navier-)Stokes bilinear form. In the presented computations the grad-div parameter was chosen as $\gamma_{gd}=0.1$. For the convective term we use the standard convective form. The computations are performed using the high-order finite element library \texttt{Netgen}/\texttt{NGSolve} \cite{Schoeberl97,Schoeberl14}. Local element unknowns of higher-order elements are eliminated from the global system via a Schur-compliment and the local mesh size on the obstacle $\nicefrac{h_{max}}{5}$ in both two and three dimensional computations. In the two dimensional case we chose $k=5$ and in three dimensions $k=4$.

\subsubsection{Details for the stationary computations}\label{SubSec:NumHvWstationary}

We use a Newton approach to directly compute the steady state solution in order to solve the problems \textbf{Rot2d-1} and \textbf{Rot3d-1}. The angular velocity is identified by using a Newton method so solve the problem
\begin{equation*}
	j(\omega) \coloneqq \Tb(\ub(\omega)) \stackrel{!}{=} 0.
\end{equation*}  
The Jacobian $j'(\omega)(\delta\omega)$ is computed using a finite difference approximation 
\begin{equation*}
	j'(\omega)(\delta\omega)\approx\frac{\Tb(\ub(\omega+\delta))-\Tb(\ub(\omega))}{\delta} .
\end{equation*} 
For our computations we take $\delta=10^{-4}$. This approximation of the Jacobian is only updated if the torque has not sufficiently decreased with the update of the angular velocity. The arising non-symmetric linear systems are solved using the sparse direct solver \texttt{Intel MKL PARDISO} \cite{IntelMKL}. The functionals for drag lift and torque are evaluated using the Babu\v{s}ka-Miller trick, cf. Babu\v{s}ka and Miller \cite{BabuskaMiller1984a} or Richter\cite[Remark 8.17]{Richter2017}.

\subsubsection{Details on the non-stationary computations}\label{SubSec:NumHvWunsteady}

To decouple the Navier-Stokes equations \eqref{eqn:NavierStokes} from the rotational ODE \eqref{eqn:RotationODE} in the system \eqref{eqn:FluidBoundaryCond}, we use a (semi-) implicit scheme to solve the PDE and use an explicit scheme to solve the ODE.

For the temporal discretisation of the grad-div stabilised Taylor-Hood method we use the second order SBDF2 Implicit-Explicit (IMEX) time-stepping scheme\cite{AscherRuuthWetton1995}. This scheme uses the BDF2 scheme for the discretisation of the time derivative and the linear Stokes part while a second order extrapolation is used to treat the convective term explicitly. As a result, we solve the same linear system in each time-step with changing right-hand sides. We denote this system as $M^\ast$. In this system we use the standard weak form of the Stokes term
\begin{equation*}
	\nu(\nabla\ub_h,\nabla\vb_h) - (p_h,\div(\ub_h)) - (q_h,\div(\ub_h))
\end{equation*}
rather than the full symmetric stress tensor which would result in the viscous term $\nu(\nabla\ub_h + \nabla\ub_h^T,\nabla\vb_h)$. As a result of this, the outflow condition in \eqref{eqn:FluidBoundaryCond} is the natural do-nothing condition, meaning that we do not have to correct for terms on $\Gout$, which would otherwise appear from integration by parts of \eqref{eqn:NavierStokes}. This has the effect that the system $M^\ast$ we have to solve in each time step is symmetric, which we take advantage of. The resulting linear system is solved using NGSolve's sparse direct solver \texttt{sparsecholesky}. Nevertheless, the full symmetric Cauchy stress tensor is used to compute the forces acting on the obstacle. Rather than computing \eqref{eqn:DefTorque} and \eqref{eqn:DirectionalForces} directly, we us an equivalent volume integral formulation to compute the forces, since this is more stable and accurate than the boundary integral formulation\cite{John2004}. We further note that the explicit treatment of the convective term results in a CFL condition, restricting the size of the time-step.

In conjunction with this IMEX scheme, we use the explicit part of the schemes to advance the angular velocity, i.e. we use the BDF2 formula to discretise the time derivative and use a second order extrapolation for the right-hand side.

\subsection{\texorpdfstring{High-order, exactly divergence-free hybrid discontinuous Galerkin method (HDG\textsubscript{$\bm{H^{div}}$})}{
	High-order, exactly divergence-free hybrid discontinuous Galerkin method}}
	
As another method, we consider a space discretisation using high-order, exactly divergence-free hybrid discontinuous Galerkin (HDG) method based on $H(\operatorname{div})$-conforming finite elements as presented in \cite{LehrenfeldSchoeberl16}. In this discretisation an \emph{only} $H(\operatorname{div})$-conforming finite element space $\Sigma_h$ is used for the discretization of the velocity field, i.e. BDM elements on simplices and Raviart-Thomas elements on quads. The pressure space $Q_h$ is chosen as $Q_h = \operatorname{div} \Sigma_h$ which renders the weak divergence constraint a strong divergence constraint yielding solenoidal discrete solutions.
Elements in $\Sigma_h$ are only normal-continuous and especially not $H^1$-conforming. To incorporate tangential continuity techniques from Discontinuous Galerkin (DG) methods are required. To eliviate the costs of these methods, hybrid versions of DG methods are applied here. In hybrid DG methods additional polynomial unknowns on the facets -- here polynomials in tangential direction only -- are introduced. These unknowns are used to break up the direct direct communication between element neighbours and allowing for static condensation.

The resulting spatial discretisation has several benefits such as energy stability and pressure robustness \cite{JLMNR17,SLLL_SEMA_2018} while allowing for efficient and high order accurate implementations \cite{LehrenfeldSchoeberl16,SJLLLS_CAMWA_2019}. 

The discretisation of the viscous term is based on a hybrid version of the well-known symmetric interior penalty method (with stabilisation parameter $\alpha=20$, cf. \cite{L_MTH_2010} for an introduction to HDG methods) whereas a standard Upwind technique is applied for the discretisation of the convection.

For the two-dimensional problems we use a triangular mesh with mesh size $h=0.1\cdot 2^{-L}$, for $L \in \{0,1,..,5 \}$ with two layers of anisotropic quadrilaterals in the region with distance $\min\{0.02,h\}$ around the circle. Around the obstacle elements are curved up to the order $k$ that is also used for the velocity space. For the three-dimensional problems we take the coarsest mesh which was also used for the computations in Section \ref{SubSec:NumHvWstationary}.
For convergence studies we consider mesh refinements and refinements of the polynomial degree resulting in different pairs of mesh levels and polynomial orders $(L,k)$.
All the examples in this study with the $H(\operatorname{div})$-conforming finite element method are computed with the
high-order finite element library \texttt{Netgen}/\texttt{NGSolve} \cite{Schoeberl97,Schoeberl14}.

\subsubsection{Details for the stationary computations}
For the stationary benchmarks problems \textbf{Rot2d-1} and \textbf{Rot3d-1} we use a Newton's method with a sparse direct solver \texttt{Intel MKL PARDISO} \cite{IntelMKL} for the linear systems as in Section \ref{SubSec:NumHvWstationary}. The Jacobi matrix is reused several times until the residual reduction becomes insufficient. As starting values for the Newton iteration we use the solution of the Stokes problem.

\subsubsection{Details for the unsteady computations}
Again, we use the same approach as in Section \ref{SubSec:NumHvWunsteady}, i.e. an operator splitted time integration with the same SBDF2 scheme. Let us note that due to the explicit treatment the Upwinding of the convection discretisation does not enter the linear systems that need to be solved. As a consequence the linear systems that need to be solved for are symmetric which allows to apply superconvergence strategies from HDG methods, cf. \cite[Section 2.2]{LehrenfeldSchoeberl16}. Here, we compute the functionals \eqref{eqn:DefTorque} and \eqref{eqn:DirectionalForces} directly from boundary integrals. For the time step size we use the heuristic $\Delta t = \Big(\Big\lceil 1250 \cdot 2^L \cdot \left(\frac{k}{5}\right)^{3/2} \Big\rceil\Big)^{-1}$.
For accessing the minima and maxima of the time dependent quantities of interest, we use a cubic spline interpolation of the equidistantly sampled data obtain after every time step.

\subsection{\texorpdfstring{Newton-multigrid for equal order finite elements with
  local projection stabilisation (EO\textsubscript{LPS})}{Newton-multigrid for 
  equal order finite elements with local projection stabilisation}}

Space discretisation is based on biquadratic (triquadratic in 3d)
finite elements on quadrilateral and hexahedral meshes for velocity
and pressure unknowns. To stabilise the inf-sup condition a local
projection method based penalisation of the divergence condition with
respect to the space of bilinear finite elements is employed, i.e. we add
\begin{equation*}
	\sum_{K\in\Omega_h} \alpha_K \big(\nabla(p_h-i_{h}^{(1)} p_h),\nabla (q_h-i_{h}^{(1)} q_h)\big)_T
\end{equation*}
to the weak discrete Navier-Stokes equation. By $i_h^{(1)}:V_h^{(2)}\to V_h^{(1)}$ we denote the discrete interpolation from the space of biquadratic functions $V_h^{(2)}$ to the space of bilinear functions $V_h^{(1)}$ on the same mesh $\Omega_h$.  
The stabilisation parameter is chosen as
$\alpha_K=0.1h_K \cdot (\nu h_K^{-1} + \|\ub_h\|_{\infty,K})^{-1}$ and depends on the local mesh size $h_K$ on element $K\in\Omega_h$ and the local  
velocity. Details are given in~\cite{BeckerBraack2001}. 
For temporal discretisation of the Navier Stokes equations and the
rigid body problem we use BDF methods of order 3. Surface
integrals like drag or torque are evaluated in variational
formulation based on the Babu\v{s}ka-Miller trick,
see~\cite{BabuskaMiller1984a} or~\cite[Remark
  8.17]{Richter2017}. Given sufficient regularity of the solution we
obtain fourth order convergence in these functional
values~\cite{BraackRichter2006d}.

The arising algebraic systems are approximated by Newton's method,
where the Jacobian is computed analytically. Linear systems of
equations are solved by a GMRES method, preconditioned with a
geometric multigrid solver. As smoother we either employ an incomplete
lower upper decomposition or a block-Jacobi iteration to employ some
parallelisation. Details are described
in~\cite{FailerRichter2019,KimmritzRichter2010}. 

\subsubsection{Details on the stationary benchmark problems}

The approach as described in Section~\ref{SubSec:NumHvWstationary} is
used to compute the stationary states in \textbf{Rot2d-1} and
\textbf{Rot3d-1} directly. For the finite difference approximation of
the Jacobian, $\delta=10^{-3}$ is taken. An alternative to determine 
the Jacobian is to solve tangent problems, however, we do not take this 
approach to avoid costly matrix assemblies that would be required. For a finite difference approximation the available Jacobian can be reused. 

\subsubsection{Details on the non-stationary benchmark problems}

Implicit discretisation methods (BDF-3) are used for the
Navier-Stokes equations and the rigid body problem. Coupling of the
two problems is achieved by a simple outer iteration. A first
prediction for the angular velocity is obtained by the explicit second
order Heun's  formula. In every time step $t_n\to t_{n+1}$ we
first predict the angular velocity
  $\omega_n\xrightarrow[(\ub_n,p_n)]{} \omega_{n+1}^{(0)}$ and set
$\ub_{n+1}^{(0)}=\ub_n$. For $l=1,2,\dots$ we iterate
\begin{enumerate}
\item Solve the Navier-Stokes equations $\ub_{n+1}^{(l-1)}
  \xrightarrow[\omega_{n+1}^{(l-1)}]{}
  (\ub_{n+1}^{(l)},p_{n+1}^{(l)})$ with BDF-3
\item Solve the rigid body problem $\omega_{n+1}^{(l-1)}
  \xrightarrow[(\ub_{n+1}^{(l)},\ub_{n}^{(l)})]{}
  \omega_{n+1}^{(l),*}$ with BDF-3
  and relax the update
  $\omega_{n+1}^{(l)}:=(1-\gamma)\omega_{n+1}^{(l-1)}+\gamma\omega_{n+1}^{(l),*}$ 
\item Stop if $|\omega_{n+1}^{(l)}-\omega_{n+1}^{(l-1)}|\le
  10^{-10}\cdot k$ where $k$ is the time step. 
\end{enumerate}
The relaxation parameter is mostly set to $\gamma=1$. Usually only two
or three iterations are required and in most cases, the Navier-Stokes
equations have to be solved only once as the initial Newton residual
is still below the tolerance after updating $\omega_{n+1}^{(l)}$. 

Since implicit time stepping formulas allow for large time step sizes,
the periodic \textbf{Rot2d-2} benchmark problem requires a good
capturing of the period length $\nicefrac{1}{f}$. After each
approximate period $t_0\mapsto t_n$ we therefore update the time-step
to minimise
\[
\sum_{i=1}^{n-1} |{\cal P}^2[\omega](t_i+s)-\omega(t_{i-n})|^2\to \min,
\]
where ${\cal P}^2[\omega]$ is a quadratic reconstruction of the discrete
angular velocities. Expecting fourth order convergence in space, we
combine each mesh refinement $h \mapsto 2^{-1}\cdot h$ with a time
step refinement of $k\mapsto 2^{-\nicefrac{4}{3}} k$ to globally
achieve fourth order. 

\subsection{Standard Taylor–Hood elements (TH)}

We consider standard Taylor–Hood finite elements\cite{Taylor1973} consisting of continuous $\Poly^2$ Lagrange elements for the velocity component and continuous $\Poly^1$ Lagrange elements for the pressure field. As mentioned above this pair is inf-sup stable.

The implementation is done in \texttt{FEniCS (2018.1.0)} \cite{FenicsBook,FenicsVersion}. The simulation are run on the prebuilt Anaconda Python packages from \cite{fenicsweb}. In 2D, the meshes where made using the built-in FEniCS mesh generation component \texttt{mshr}  and the circular obstacle was resolved with a local mesh size $\nicefrac{h_{max}}{8}$. In 3D the meshes where made using \texttt{Netgen}\cite{Schoeberl97} with the local mesh size at the sphere also of $\nicefrac{h_{max}}{8}$ and converted with a \texttt{Python} script \texttt{dolfin-convert}.

The forces are computed based on the equivalent volume integral formulation which is more accurate than the boundary formulation as well as more stable and robust with respect to the boundary approximation \cite{John2004}.

\subsubsection{Details on the stationary benchmark problem}

Stationary computations follow the approach described in Section~\ref{SubSec:NumHvWstationary}. However, the parallel sparse direct solver MUMPS\cite{mumps} was used to solve the arising linear systems.

\subsubsection{Details on the non-stationary benchmark problems}

Temporal discretisation of the Navier-Stokes system \eqref{eqn:NavierStokes} is realised using the second order Crank-Nicolson scheme. The resulting non-linear systems are then solved using Newtons method. For time discretisation of the ordinary differential equation for the angular velocity \eqref{eqn:RotationODE} we employ the appropriate explicit linear multistep scheme, the second order Adams-Bashforth method.

\subsection{\texorpdfstring{Taylor-Hood and Optimisation or Method of Lines (TH\textsubscript{\texttt{scipy}}) }{
	Taylor-Hood and Optimisation or Method of Lines}}

The spatial discretisation was done with Taylor-Hood $\Poly^2/\Poly^1$ elements in \texttt{FEniCS} \cite{FenicsBook} via its \texttt{Python} interface \texttt{dolfin}. With the help of the \texttt{Python} module \texttt{dolfin\_navier\_scipy} \cite{swHei19}, all discrete operators were interfaced to \texttt{SciPy}\cite{scipy} for the solution of the steady-state problem, the optimisation, and the numerical time integration. 

The Dirichlet boundary values were included by directly assigning the relevant values to the corresponding \emph{degrees of freedom}; see \cite[Rem. 3.2]{BenH15} on how to adjust a time stepping scheme accordingly.

The boundary integrals in the functionals for drag lift and torque were approximated using the Babu\v{s}ka-Miller trick, cf. \cite{BabuskaMiller1984a} or \cite[Remark 8.17]{Richter2017}.

\subsubsection{Details on the stationary benchmark problem}\label{SubSec:NumJHstationary}

As mentioned in Section~\ref{SubSec:NumHvWstationary}, the stationary limit can be computed via finding the $\omega$ for which the resulting torque force is zero:
\begin{equation}
\Tb(\ub(\omega)) \stackrel{!}{=} 0.
\end{equation}

We use the builtin function \texttt{scipy.optimize.brent} which employs the \emph{brent} algorithm \cite{Brent71} to minimise the scalar univariate function $\omega \mapsto \Tb(\ub(\omega))^2$.

\subsubsection{Details on the non-stationary benchmark problem}

For the time integration, we use the Crank-Nicolson scheme on the linear part of the momentum equation and second order Adams-Bashforth for the nonlinear part, while enforcing the incompressibility constraint explicitly in every time step. This implict-explicit scheme is of order two and was already used in \cite{KimP85} and analysed in \cite{MarT98}. Because of the explicit treatment of the convection and because of the equidistant time grid, in every time step, the same linear system is to be solved. For this, an \emph{LU factorisation} of the corresponding coefficient matrix was precomputed. The ODE for the angular velocity \eqref{eqn:RotationODE} was discretised with the second order Adams-Bashforth method.

To reduce the time until the system has reached the limit cycle, we started from the corresponding steady-state solution with angular velocity $\omega$ such that $\Tb(\ub(\omega)) \approx 0$ as computed with the optimisation method described above in Section~\ref{SubSec:NumJHstationary}. For higher Reynolds numbers the Newton iterations for the steady state converge only locally. We obtained convergence with the steady states solutions that correspond to $\nu=800$ or $Re = 80$ as initial guess.

%% file: results.tex
\section{Results}\label{sec:Results}

We present the numerical results for the benchmark problems obtained by the various methods described in Section~\ref{sec:NumMet}. Next to the reference values, we also provide further information on the discretisation through the number of degrees of freedom and the number on non-zero entries of the resulting linear systems which needed to be solved. The non-zero entries indicate the sparsity of the resulting matrix and therefore the effort needed to solving the system.

For TH\textsubscript{gd,ho} and HDG\textsubscript{$H^{div}$} we also provide the number of degrees of freedom after removing internal degrees of freedom from the global system using a Schur complement, known as static condensation. As a result the non-zero entries are also only counted for the system after static condensation, i.e., the System which has to be solved in the method. Furthermore, we provide hardware information and computation times for the simulations. This is simply to indicate how challenging the proposed benchmark problems are.

\paragraph{Rot-2d1}
The results obtained for the stationary \emph{Rot-2d1} benchmark, computed using the methods described in Section~\ref{sec:NumMet}, can be seen in Table~\ref{tab:ResultsRot2D1}. We summarise these results in Table~\ref{tab:RefValsRot2D1}, where we give reference intervals for the quantities of interest based on the full results in Table~\ref{tab:ResultsRot2D1}.

We see that the reverence intervals have sizes between the order of magnitude $10^{-5}$ for $C_D$ and $10^{-8}$ for $\omega^\ast$. This in combination with the relatively short computation times shows that the problem \emph{Rot-2d1} can be seen as easily computable. Furthermore, we see that higher-order methods are particularly suited to this problem and that the \emph{direct} quasi Newton approach as described in Section~\ref{SubSec:NumHvWstationary} is particularly efficient for this problem.

\paragraph{Rot-2d2}

\begin{figure}
	\begin{minipage}[t]{0.47\textwidth}
		\input{plot2d2}
	\end{minipage}
	\hspace{0.05\textwidth}
	\begin{minipage}[t]{0.47\textwidth}
		\input{plot2d3}
	\end{minipage}
	\caption{The quantities of interest for the problem Rot-2d2 over two periods (left) and the quantities of interest for the problem Rot-2d3 over the entire time-interval (right).}
	\label{fig:Rot2D2-3-plots}
\end{figure}
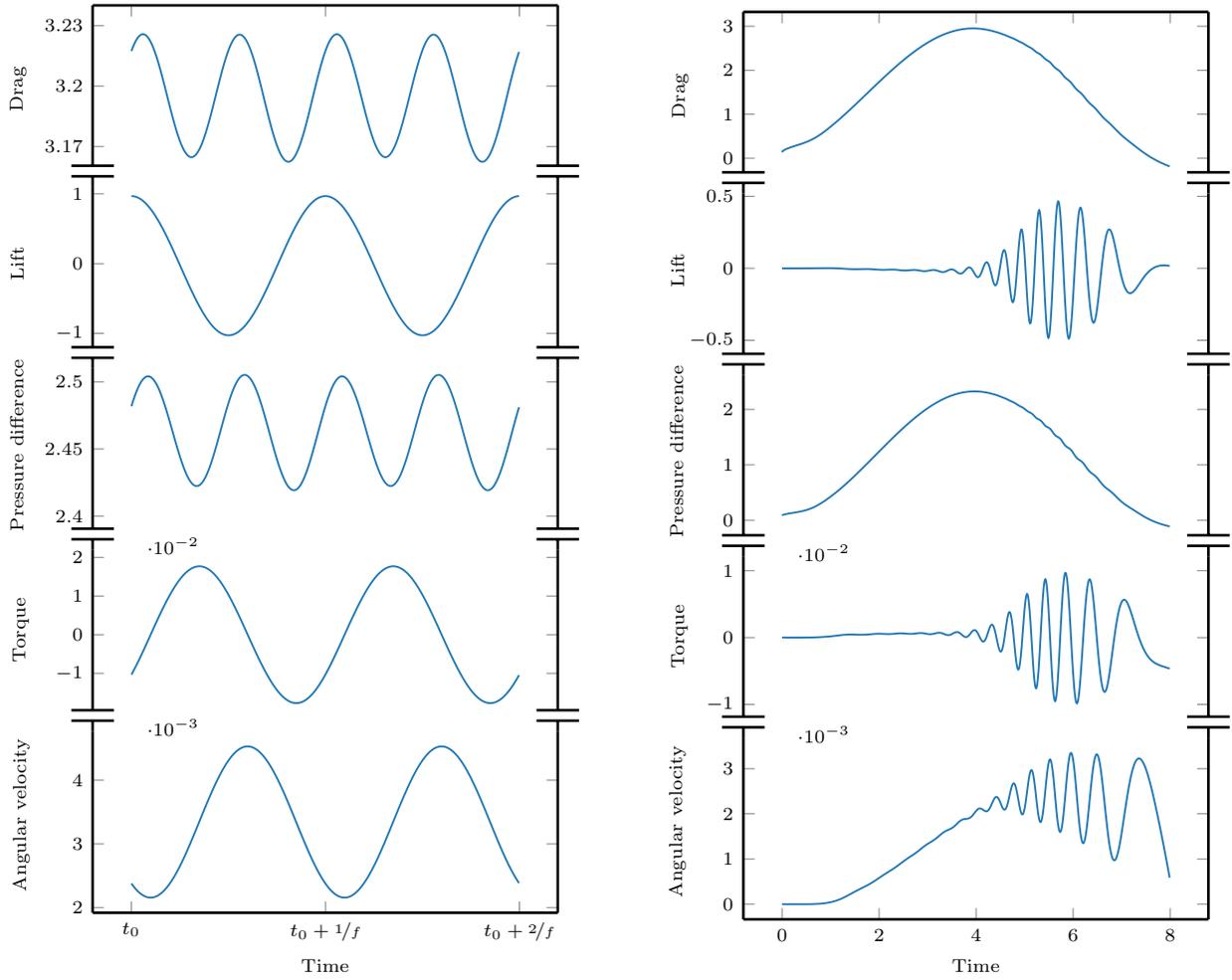

The results for the problem \emph{Rot-2d2} are summarised in Table~\ref{tab:ResultsRot2D2}. Ranges for the reference values for the quantities of interest for this problem are given in Table~\ref{tab:RefValsRot2D2}. A plot of the quantities of interest over two periods can be seen in Figure~\ref{fig:Rot2D2-3-plots}.

We can see here that the non-stationary problem \emph{Rot-2d2} is more challenging, because the interval lengths for the reference values is between the order $10^{-3}$ and the order $10^{-5}$. We also note that the angular velocity seem to be comparably easy quantities to compute as the reference intervals are smallest for this quantity.

\paragraph{Rot-2d3}

The quantities of interest in the problem \emph{Rot-2d3} are presented in Table~\ref{tab:ResultsRot2D3} and the resulting reference intervals are given in Table~\ref{tab:RefValsRot2D3}. The quantities of interest are also plotted over the entire time interval in Figure~\ref{fig:Rot2D2-3-plots}. 

With respect to the spatial discretisation, high-order methods have performed very well here. With respect to the temporal integration, we note that this problem seems to be challenging. On the one hand, the forth order (by the choice of time-steps) time-discretisation in TR gave good results with relatively large time steps while the second order SBDF2 IMEX time stepping with TH\textsubscript{gd,ho} also gave good results on (relatively) coarser meshes in combination with very small time-steps. This problem therefore seems particularly suited to assess the performance of time-integration methods for Navier-Stokes.

With respect to accuracy of the individual values, we note the most difficult to compute value with respect to the absolute error is $C_{L,\text{max}}$ while we can give the smallest reference interval for $\omega^\ast_{\text{max}}$. However, the times at which the extreme values are realised are all only accurate up to the order $10^{-3}$.

Looking at the computational times we note that the problem is significantly more challenging than the stationary problem, however, it is also still a reasonable problem to compute. 

\paragraph{Rot-3d1}

The values computed for the problem Rot-3d1 can be seen in Table~\ref{tab:ResultsRot3D1}. We summarise these results in reference intervals given in Table~\ref{tab:RefValsRot3D1}.

From the reference intervals we see that $\omega^\ast_1$ is especially challenging to compute accurately since it is relatively small in comparison to the other values. The change of sign between different computations is therefore not surprising. The force coefficient $C_{\Fb_2}$ seems to be the easiest to compute while the remaining values seem equally challenging. 

\begin{table}
  \centering
    \resizebox{\textwidth}{!}{%
      \begin{tabular}{crr llrll r}
        \toprule
        \multicolumn{3}{c}{Discretisation} & \multicolumn{6}{c}{Results}\\
        \cmidrule(lr){1-3}\cmidrule(lr){4-9}
        Method	&	\#dof [$K$]	&	\#nze [$M$]	&	$C_D$	&	$C_L$ &	\multicolumn{1}{l}{$C_T$}	&	$\Delta p$	&	$\omega^\ast$	& Comp.T. [$s$]	\\
        \midrule
		TH\textsubscript{gd,ho}\footnotemark[1]	& 15.6(8.0)		& 0.6	& 5.5795523091	& 0.0047151364	& -4.10e-16	& 0.1174948237	& 0.0012629527	& 4.1 \\
		$k=5$				& 51.1(26.8)		& 2.0	& 5.5795587061	& 0.0047141285	& -3.03e-16	& 0.1175213881	& 0.0012629334	& 6.2 \\
		 					& 186.5(99.7)	& 7.5	& 5.5795588114	& 0.0047141929	& 6.03e-16	& 0.1175205313	& 0.0012629345	& 19.6 \\
		 					& 693.6(374.2)	& 28.0	& 5.5795588133	& 0.0047141930	& 7.28e-16	& 0.1175202728	& 0.0012629346	& 68.1 \\
		\cmidrule(lr){1-9}
  		EO\textsubscript{LPS}\footnotemark[2]\textsuperscript{,}\footnotemark[3]%
			&33.4 	& 0.5	&5.579526747	&0.004727828	&1.67e-12	&0.117518677	&0.00126296256	&5.7\\
			&131.9	& 2.1	&5.579556532	&0.004715150	&-2.15e-15  &0.117519524	&0.00126292998	&31.6\\
			&523.9 	& 8.3	&5.579558664	&0.004714255	&-2.17e-15  &0.117520022	&0.00126293418	&137.2\\
			&2088.1	& 33.3	&5.579558804	&0.004714197
                &-4.47e-16  &0.117520193 	&0.00126293459
                &604.3\\
		\cmidrule(lr){2-9}
		&\multicolumn{2}{l}{extrapolated}&
                5.579558814&0.004714193&- & 0.117520282&0.00126293463 \\
		&\multicolumn{2}{l}{(order)}& (3.93) & (3.95) & - &(2.00)
                & (3.36)\\
 		\cmidrule(lr){1-9}
        TH\footnotemark[4]
		&  9.8 & 0.3  &  5.5908135746  & 0.0034359338 & 0 & 0.1175346984 & 0.0011602113 &  0.4 \\ 
		&  29.7 & 0.8  &  5.5829138831  & 0.0054926020 & -2.71e-16 & 0.1174672688 & 0.0014131732 &  1.2 \\ 
		&  99.7 & 2.9  &  5.5801404856  & 0.0048778267 & 2.37e-16 & 0.1175411687 & 0.0012546495 &  4.6 \\ 
		&  362.4 & 10.6  &  5.5795742677  & 0.0047489983 & 0 & 0.1175218581 & 0.0012574215 &  20.3 \\ 
		&  1379.1 & 40.6  &  5.5795553554  & 0.0047154167 & -1.07e-13 & 0.1175158081 & 0.0012628280 &  127.0 \\ 
		&  5370.2 & 158.7  &  5.5795581671  & 0.0047141845 & 1.22e-14 & 0.1175193123 & 0.0012629381 &  657.0 \\   
		\cmidrule(lr){1-9}                                                      
		HDG\textsubscript{$H^{div}$}\footnotemark[5] 
        $(L,k)=(0,\hphantom{0}5)$ &
		8.9(5.2)	&0.4	&5.5795057887	&0.0047518775	&-9.99e-13	&0.1175238245	&0.0012598766	&6.5\\
		         \hphantom{HDG\textsubscript{$H^{div}$}}        $(L,k)=(0,\hphantom{0}6)$ &
		25.4(6.1)	&0.5	&5.5795547031	&0.0047188391	&-7.19e-14	&0.1175209652	&0.0012621510	&8.3	\\
		         \hphantom{HDG\textsubscript{$H^{div}$}}        $(L,k)=(0,\hphantom{0}7)$ &
		33.0(7.0)	&0.6	&5.5795584283	&0.0047140877	&-7.44e-13	&0.1175200945	&0.0012629298	&9.9	\\
		         \hphantom{HDG\textsubscript{$H^{div}$}}        $(L,k)=(0,\hphantom{0}8)$ &
		41.5(7.8)	&0.8	&5.5795588381	&0.0047141851	&-7.81e-13	&0.1175202550	&0.0012629342	&19.1	\\
		         \hphantom{HDG\textsubscript{$H^{div}$}}        $(L,k)=(0,\hphantom{0}9)$ &
		51.0(8.7)	&1.0	&5.5795588122	&0.0047141915	&-9.51e-16	&0.1175202619	&0.0012629337	&16.3	\\
		         \hphantom{HDG\textsubscript{$H^{div}$}}        $(L,k)=(0,10)$ &
		61.5(9.6)	&1.2	&5.5795588016	&0.0047141905	&-3.12e-13	&0.1175202597	&0.0012629346	&67.5   \\
		\cmidrule(lr){1-9}                             
		HDG\textsubscript{$H^{div}$}\footnotemark[5]         $(L,k)=(0,\hphantom{0}5)$ &
		18.9(5.2)	&0.4	&5.5795057887	&0.0047518775	&-9.99e-13	&0.1175238245	&0.0012598766	&6.5	\\
		         \hphantom{HDG\textsubscript{$H^{div}$}}        $(L,k)=(1,\hphantom{0}5)$ &
		50.1(14.7)	&0.9	&5.5795397082	&0.0046822197	&-8.23e-13	&0.1175243136	&0.0012676502	&12.5	\\
		         \hphantom{HDG\textsubscript{$H^{div}$}}        $(L,k)=(2,\hphantom{0}5)$ &
		181.1(55.4)	&3.4	&5.5795402179	&0.0047150139	&7.35e-13	&0.1175244061	&0.0012628701	&23.4	\\
		         \hphantom{HDG\textsubscript{$H^{div}$}}        $(L,k)=(3,\hphantom{0}5)$ &
		745.5(231.8)	&14.1	&5.5795566498	&0.0047142079	&2.89e-14	&0.1175205506	&0.0012629366	&126.0	\\
		         \hphantom{HDG\textsubscript{$H^{div}$}}        $(L,k)=(4,\hphantom{0}5)$ & 
		2949.9(924.2)	&55.8	&5.5795587277	&0.0047141940	&-7.16e-13	&0.1175202662	&0.0012629347	&615.3	\\		
		\cmidrule(lr){1-9}                             
		 TH\textsubscript{\texttt{scipy}}\footnotemark[6] 
		 & 30.9 & 0.6 & 5.5787704610 & 0.0047091588 & -1.84e-13 & 0.1174850027 & 0.0012628449 & 68.1 \\
		 & 77.9 & 1.6 & 5.5792205094 & 0.0047130497 & -2.29e-13 & 0.1175153993 & 0.0012627522 & 284.2 \\
		 & 273.0 & 5.7 & 5.5794655316 & 0.0047136722 & -2.04e-13 & 0.1175255316 & 0.0012628650 & 2668.0 \\
		 & 696.4 & 14.5 & 5.5795218195 & 0.0047139943 & -2.90e-14 & 0.1175224057 & 0.0012629083 & 13806.5 \\
		 & 1931.4 & 40.2 & 5.5795454842 & 0.0047141221 & -1.32e-12 & 0.1175200148 & 0.0012629253 & 68856.1 \\
        \bottomrule        
      \end{tabular}
    }
    \caption{Results for Rot-2D1. The number of degrees of freedom (dof) is given in $K=10^3$. Dofs in brackets are the number of unconstrained dofs after static condensation. The number of non-zero entries (nze) of the condensed (if applicable) system matrix are given in $M=10^6$. The computational run-time is given in seconds $s$.}
	\label{tab:ResultsRot2D1}
\end{table} 
\footnotetext[1]{Computed using AMD Ryzen 5 2400G, 3.6GHz using 4 cores.}
\footnotetext[2]{Computed on MacBookPro using Intel i7-4870HQ, 2.5 GHz using 4 cores.}
\footnotetext[3]{Extrapolated values are indicated if fitting to $a(h) = a+ch^q$ is possible.}
\footnotetext[4]{Computed using Intel Core E5620, 2.4GHz with 4 cores and 8 threads.}
\footnotetext[5]{Computed using Intel Core i7-8650U, 4.2GHz with 4 cores and 8 threads.}
\footnotetext[6]{Computed using Intel Xeon E7-8837, 2.67GHz, using 1 (physical and virtual) core.}

\begin{table}
	\centering
	\resizebox{\textwidth}{!}{%
	\begin{tabular}{r cccc}
		\toprule
		Coefficient & $C_D$		& $C_L$		& $\Delta p$	& $\omega^\ast$\\
			\cmidrule(lr){2-5}
		Value range		& $[5.57954,5.579559]$ & $[0.0047141,0.0047142]$ & $[0.117519,0.117521]$ 	& $[0.001262925,0.00126294]$ \\
		\bottomrule
	\end{tabular}
	}
	\caption{Reference value ranges for the quantities of interest for the problem Rot-2d1.}
	\label{tab:RefValsRot2D1}
\end{table}


\begin{table}
  \centering
    \resizebox{\textwidth}{!}{%
      \begin{tabular}{crrl lllllllllll}
        \toprule
        \multicolumn{4}{c}{Discretisation} & \multicolumn{10}{c}{Results}\\
        \cmidrule(lr){1-4}\cmidrule(lr){5-14}
        Method	&	\#dof [$K$]	&	\#nze [$M$]	&	$\tau$	&	$C_{D,\text{max}}$	&	$C_{D,\text{min}}$	&	$C_{L,\text{max}}$	&	$C_{L,\text{min}}$	& $C_{T,\text{max}}$	& $C_{T,\text{min}}$	&	$\Delta p(t^\ast)$	&	$\omega^\ast_{\text{max}}$	&	$\omega^\ast_{\text{min}}$	&	$St$\\
        \midrule
		TH\textsubscript{gd,ho}		
				& 15.6(8.0) 	& 0.3 & $\nicefrac{1}{2500}$ 	& 3.2257143516	& 3.1623995194	& 0.9682189506	&-1.0272488814	& 0.0177090000	& -0.0177300000	& 2.4832272754	& 0.0045764494	 & 0.0021991719	 & 0.3019323671 \\
		$k=5$ 	& 51.1(26.8) 	& 1.1 &	$\nicefrac{1}{5000}$ 	& 3.2257765946	& 3.1624660264	& 0.9676789054	&-1.0273532513	& 0.0176960000	& -0.0177170000	& 2.4834909048	& 0.0045289959	 & 0.0021540270	 & 0.3019323671 \\
				& 186.5(99.7) 	& 3.9 & $\nicefrac{1}{10000}$	& 3.2257012749	& 3.1624171702	& 0.9674937697	&-1.0271414624	& 0.0176920000	& -0.0177140000	& 2.4840487990	& 0.0045308702	 & 0.0021563680	 & 0.3018412315 \\
				& 693.6(374.2) 	& 14.5 & $\nicefrac{1}{20000}$	& 3.2257003973	& 3.1624161918	& 0.9674952132	&-1.0271424650	& 0.0176920000	& -0.0177140000	& 2.4841879091	& 0.0045307692	 & 0.0021562635	 & 0.3018412315 \\
		\cmidrule(lr){1-14} 
        EO\textsubscript{LPS}\footnotemark[3]
		&22.5&1.0&$\nicefrac{t_P}{20}$& 3.265238       & 3.184170 & 1.092590 & -1.152795 & 0.0199021 & -0.01992426 &2.516449& 0.00468163 & 0.00199515&0.299292 \\
		&88.0&4.1&$\nicefrac{t_P}{50}$& 3.229858       & 3.164937 & 0.980084 & -1.039833 & 0.0179042 & -0.01792674 &2.487680& 0.00454846 & 0.00214495&0.301792 \\
		&348.0&16.5&$\nicefrac{t_P}{126}$& 3.225959    & 3.162583 & 0.968258 & -1.027915 & 0.0177047 & -0.01772654 &2.484521& 0.00453230 & 0.00215602&0.301871 \\
		&1\,348.1&66.1&$\nicefrac{t_P}{320}$& 3.225713 & 3.162425 & 0.967522 & -1.027180 & 0.0176923 & -0.01771480 &2.484473& 0.00453212 & 0.00215744&0.301869 \\
		\cmidrule(lr){2-14}
		&\multicolumn{3}{l}{extrapolated}& 3.225697 & 3.162414 & 0.967473 &
		-1.027131 & 0.0176915 & -0.01771407 &2.484318& 0.00453120 & 0.00215765& c.o.s.\\
		&\multicolumn{3}{l}{(order)}& (4.00) & (3.90) & (4.01) & (4.02) & (4.01)&(4.09)&(3.27)& (3.14) & (2.96) & -\\
                \cmidrule(lr){1-14} 
		TH
		& 9.8 & 0.3 & \nicefrac{1}{40}	& 3.1140915806	& 3.0835435234	& 0.6942111384	&-0.7604045006	& 0.0140683780	& -0.0140959295	& 2.4312708845	& 0.0055445513	 & 0.0034041747	 & 0.2884975288\\
		& 29.7 & 0.8 & \nicefrac{1}{80}	& 3.2150998204	& 3.1562985910	& 0.9535596576	&-1.0168443659	& 0.0176195464	& -0.0176358975	& 2.4690328925	& 0.0048579972	 & 0.0024203675	 & 0.2995190345\\
		& 99.7 & 2.9 & \nicefrac{1}{160}	& 3.2245565079	& 3.1614311213	& 0.9674861840	&-1.0274123288	& 0.0176662978	& -0.0176716112	& 2.4739243898	& 0.0045337913	 & 0.0021466233	 & 0.3014554428\\
		& 362.4 & 10.6 & \nicefrac{1}{320}	& 3.2256301394	& 3.1621590399	& 0.9687958092	&-1.0285405038	& 0.0178384619	& -0.0178564492	& 2.4826673308	& 0.0045427458	 & 0.0021447040	 & 0.3017710165\\
		& 1379.1 & 40.6 & \nicefrac{1}{640}	& 3.2256835394	& 3.1622852649	& 0.9685324763	&-1.0281714901	& 0.0178273180	& -0.0178503246	& 2.4837365764	& 0.0045386903	 & 0.0021448458	 & 0.3018441883\\
		\cmidrule(lr){1-14} 
		TH\textsubscript{\texttt{scipy}}
		& 30.9 & 0.6 & $2^{ -12 }$ & 3.2258901997 & 3.1644408460 & 0.9719117446 & -1.0316255447 & 0.0177891954 & -0.0178119760 & 2.4836587332 & 0.0045491972 & 0.0021580693 & 0.3014232379 \\
		& 77.9 & 1.6 & $2^{ -12 }$ & 3.2253415537 & 3.1645119141 & 0.9678401829 & -1.0274808411 & 0.0177006474 & -0.0177221160 & 2.4835061003 & 0.0045297258 & 0.0021538441 & 0.3018393315 \\
		& 145.4 & 3.0 & $2^{ -13 }$ & 3.2255992838 & 3.1623146862 & 0.9675287323 & -1.0271645858 & 0.0176942199 & -0.0177156522 & 2.4839426827 & 0.0045294074 & 0.0021545924 & 0.3018633609 \\
		& 273.0 & 5.7 & $2^{ -13 }$ & 3.2256469196 & 3.1646546774 & 0.9674822148 & -1.0271283545 & 0.0176931937 & -0.0177146006 & 2.4839169565 & 0.0045306268 & 0.0021560077 & 0.3018705247 \\
		\cmidrule(lr){1-14} 
        HDG\textsubscript{$H^{div}$}      $(L,k)=(0,\hphantom{0}5)$
		&19.1	&0.2 & $\nicefrac{1}{1250}$	&
		3.2250894217	& 3.1631970522	& 0.9554878894	&-1.0167407708	& 0.0174498580	& -0.0174774565	& 2.4821304744	& 0.0046180909	 & 0.0022830661	 & 0.3026634383
		\\         \hphantom{HDG\textsubscript{$H^{div}$}}        $(L,k)=(1,\hphantom{0}5)$
		&50.3	&0.5 & $\nicefrac{1}{2500}$	& 
		3.2255461452	& 3.1623559873	& 0.9675410784	&-1.0266478889	& 0.0176809137	& -0.0177017106	& 2.4841689387	& 0.0044381232	 & 0.0020651124	 & 0.3019323671
		\\         \hphantom{HDG\textsubscript{$H^{div}$}}        $(L,k)=(2,\hphantom{0}5)$
		&181.5	&2.0	& $\nicefrac{1}{5000}$	& 
		3.2256839724	& 3.1623930523	& 0.9675089999	&-1.0272149646	& 0.0176928223	& -0.0177140904	& 2.4841763850	& 0.0045347470	 & 0.0021601730	 & 0.3019323671
		\\         \hphantom{HDG\textsubscript{$H^{div}$}}        $(L,k)=(3,\hphantom{0}5)$
		&746.2	&8.2	& $\nicefrac{1}{10000}$	&
		3.2256935844	& 3.1624101505	& 0.9674881509	&-1.0271362492	& 0.0176921047	& -0.0177135376	& 2.4841005356	& 0.0045308638	 & 0.0021563879	 & 0.3018412315
		        \\
		\cmidrule(lr){1-14} 
        HDG\textsubscript{$H^{div}$}      $(L,k)=(0,\hphantom{0}5)$
		&19.1	&0.2	& $\nicefrac{1}{1250}$	&
		3.2250894217	& 3.1631970522	& 0.9554878894	&-1.0167407708	& 0.0174498580	& -0.0174774565	& 2.4821304744	& 0.0046180909	 & 0.0022830661	 & 0.3026634383
		\\        \hphantom{ HDG\textsubscript{$H^{div}$}}        $(L,k)=(0,\hphantom{0}6)$
		&25.7	&0.3	& $\nicefrac{1}{1644}$	& 
		3.2268543979	& 3.1635587023	& 0.9671329198	&-1.0273404878	& 0.0176726465	& -0.0176938112	& 2.4843767829	& 0.0046129318	 & 0.0022434678	 & 0.3022051913
		\\         \hphantom{HDG\textsubscript{$H^{div}$}}        $(L,k)=(0,\hphantom{0}7)$
		&33.3	&0.4	& $\nicefrac{1}{2071}$	& 
		3.2263007223	& 3.1629779600	& 0.9679161173	&-1.0276532855	& 0.0176944680	& -0.0177167087	& 2.4840834627	& 0.0045702262	 & 0.0021963030	 & 0.3018949949
		\\         \hphantom{HDG\textsubscript{$H^{div}$}}        $(L,k)=(0,\hphantom{0}8)$
		&41.9	&0.5	& $\nicefrac{1}{2530}$	& 
		3.2262827883	& 3.1628681950	& 0.9686648558	&-1.0283599930	& 0.0177079714	& -0.0177294765	& 2.4845121998	& 0.0045563934	 & 0.0021801191	 & 0.3019095781
		\\         \hphantom{HDG\textsubscript{$H^{div}$}}        $(L,k)=(0,\hphantom{0}9)$
		&51.4	&0.6	& $\nicefrac{1}{3019}$  & 
		3.2259541386	& 3.1626013428	& 0.9681203559	&-1.0277726888	& 0.0177017718	& -0.0177233126	& 2.4842577357	& 0.0045419773	 & 0.0021662651	 & 0.3019004634
		\\         \hphantom{HDG\textsubscript{$H^{div}$}}        $(L,k)=(0,10)$
		&62.0	&0.7	& $\nicefrac{1}{3536}$	& 
		3.2258239459	& 3.1625056339	& 0.9678453674	&-1.0274854052	& 0.0176975287	& -0.0177189699	& 2.4843132451	& 0.0045359375	 & 0.0021607464	 & 0.3017064517
		\\
		\bottomrule        
      \end{tabular}
    }
    \caption{Results for Rot-2D2. The number of degrees of freedom (dof) is given in $K=10^3$. Dofs in brackets are the number of unconstrained dofs after static condensation. The number of non-zero entries (nze) of the condensed (if applicable) system matrix are given in $M=10^6$. The computational run-time is given in seconds $s$. Time step size $\tau$ for TR is based on a subdivision of each period $t_P\approx 0.33$. 
    }
	\label{tab:ResultsRot2D2}
\end{table}

\begin{table}
	\centering
	\resizebox{\textwidth}{!}{%
	\begin{tabular}{r ccccc}
		\toprule
		Coefficient & $C_{D,\text{max}}$	&	$C_{D,\text{min}}$	&	$C_{L,\text{max}}$	&	$C_{L,\text{min}}$	& $C_{T,\text{max}}$\\
		\cmidrule(lr){2-6}
		Value & $[3.2256,3.22571]$ & $[3.1622,3.1647]$ & $[0.9674, 0.9686]$ & $[-1.0282, -1.0271]$ & $[0.01769,0.01783]$\\
		\midrule
		Coefficient & $C_{T,\text{min}}$	&	$\Delta p(t^\ast)$	&	$\omega^\ast_{\text{max}}$	&	$\omega^\ast_{\text{min}}$	&	$St$\\
		\cmidrule(lr){2-6}
		Value & $[-0.01786,-0.01771]$ & $[2.4837,2.4844]$ & $[0.004530,0.004539]$ & $[0.002144,0.002161]$ & $[0.30171,0.30188]$ \\
		\bottomrule	
	\end{tabular}
	}
	\caption{Reference value ranges for the quantities of interest for the problem Rot-2d2.}
	\label{tab:RefValsRot2D2}
\end{table}


\begin{table}
  \centering
    \resizebox{\textwidth}{!}{%
      \begin{tabular}{crrl lllllllll r}
        \toprule
        \multicolumn{4}{c}{Discretisation} & \multicolumn{10}{c}{Results}\\
        \cmidrule(lr){1-4}\cmidrule(lr){5-14}
        Method	&	\#dof [$K$]	&	\#nze [$M$]	&	$\tau$	&	$C_{D,\text{max}}$	&	$t_{D,\text{max}}$	&	$C_{L,\text{max}}$	&	$t_{L,\text{max}}$	& $C_{T,\text{max}}$	&	$t_{T,\text{max}}$	&	$\Delta p(8)$	&	$\omega^\ast_{\text{max}}$&	$t_{\omega,\text{max}}$	& Comp.T. [$s$]	\\
        \midrule
		TH\textsubscript{gd,ho}\footnotemark[1]	
				& 15.6(8.0) 		& 0.3	& $\nicefrac{1}{2500}$	& 2.9509895176	& 3.93640	& 0.4680735702	& 5.69320	& 0.0097303511	& 5.84240	& -0.1118627466	& 0.0033807682	& 5.95560	 & 875\\
		$k=5$	& 51.1(26.8) 	& 1.1	& $\nicefrac{1}{5000}$	& 2.9509040487	& 3.93620	& 0.4657404192	& 5.69380	& 0.0096899195	& 5.84320	& -0.1118754106	& 0.0033487422	& 5.95620	 & 5597\\
				& 186.5(99.7) 	& 3.9	& $\nicefrac{1}{10000}$	& 2.9508932851	& 3.93620	& 0.4656656345	& 5.69390	& 0.0096887696	& 5.84300	& -0.1118750991	& 0.0033484815	& 5.95630	 & 41244\\
				& 693.6(374.2) 	& 14.5	& $\nicefrac{1}{20000}$	& 2.9508912256	& 3.93620	& 0.4656663341	& 5.69390	& 0.0096887894	& 5.84300	& -0.1118751459	& 0.0033484327	& 5.95630	 & 310842\\
        \cmidrule(lr){1-14} 
		EO\textsubscript{LPS}\footnotemark[7]\textsuperscript{,}\footnotemark[3]			
		&15.6	&0.7& $\nicefrac{1}{50}$	 &2.95148758 & 3.93715 & 0.54559013 & 5.73098 & 0.01121033 & 5.88498 & -0.11005404& 0.00349536 & 6.00203 &282    \\
		&46.5	&2.1& $\nicefrac{1}{126}$ &2.95083751 & 3.93629 & 0.47297707 & 5.69516 & 0.00982570 & 5.84448 & -0.11193809& 0.00336110 & 5.95803 &2090   \\
		&154.5	&7.2& $\nicefrac{1}{330}$ &2.95093223 & 3.93620 & 0.46609980 & 5.69394 & 0.00969654 & 5.84300 & -0.11187846& 0.00334929 & 5.95639 &17418  \\
		&554.7	&26.2& $\nicefrac{1}{831}$ &2.95088595 & 3.93619 & 0.46567339 & 5.69391 & 0.00968898 & 5.84295 & -0.11187501& 0.00334838 & 5.95633 &165886 \\
			\cmidrule{2-14}
			&\multicolumn{2}{l}{extrapolated}&&\multicolumn{2}{l}{change of sign}&0.46564520 &         & 0.00968851&&-0.11187479   & 0.00334830& \\
			&\multicolumn{2}{l}{(order)}&&-&-&(4.01)     &         & (4.09)     & & (4.07)  & (3.70)&&\\
        \cmidrule(lr){1-14} 
   		TH\footnotemark[4]     
   		& 3.8 & 0.1 & \nicefrac{1}{40}	& 3.0066353678	& 3.97770	& 0.0050600950	&0.98152	& 0.0010600233	& 4.06587	& -0.1208947707	& 0.0042513467	 & 5.86591	 & 17\\
		& 9.8 & 0.3 & \nicefrac{1}{80}	& 2.9647408765	& 3.94673	& 0.1848436023	&6.86969	& 0.0046721653	& 6.37981	& -0.1022916281	& 0.0028664803	 & 7.68158	 & 50\\
		& 29.7 & 0.8 & \nicefrac{1}{160}	& 2.9520976467	& 3.94280	& 0.4552154174	&5.75891	& 0.0097338773	& 5.90943	& -0.1059880107	& 0.0028602560	 & 7.68262	 & 225\\
		& 99.7 & 2.9 & \nicefrac{1}{320}	& 2.9518102419	& 3.94043	& 0.4637777077	&5.70197	& 0.0096836137	& 5.84945	& -0.1117757539	& 0.0035836382	 & 5.95961	 & 1380\\
		& 362.4 & 10.6 & \nicefrac{1}{640}	& 2.9513456900	& 3.93838	& 0.4659134118	&5.69664	& 0.0097683463	& 5.84458	& -0.1114409580	& 0.0033407595	 & 5.95645	 & 11890\\
		& 1379.1 & 40.6 & \nicefrac{1}{1280}	& 2.9509141480	& 3.93722	& 0.4658976721	&5.69509	& 0.0097613315	& 5.84332	& -0.1116494707	& 0.0033616140	 & 5.95589	 & 117083\\
		\cmidrule(lr){1-14} 
        HDG\textsubscript{H\textsuperscript{div}}\footnotemark[5]        $(L,k)=(0,\hphantom{0}5)$
		&19.1	&0.2 & $\nicefrac{1}{1250}$	&
		2.9511386439	& 3.93680	& 0.4756634788	& 5.69280	& 0.0098269810	& 5.84240	& -0.1117569458	& 0.0034432916	& 5.95520	 &  204.9
		\\         \hphantom{HDG\textsubscript{$H^{div}$}\footnotemark[5]}        $(L,k)=(1,\hphantom{0}5)$
		&50.3	&0.5	& $\nicefrac{1}{2500}$	& 
		2.9509424810	& 3.93640	& 0.4660733580	& 5.69400	& 0.0096957993	& 5.84360	& -0.1118916796	& 0.0032765928	& 5.95680	 &  765.7
		\\         \hphantom{HDG\textsubscript{$H^{div}$}\footnotemark[5]}        $(L,k)=(2,\hphantom{0}5)$
		&181.5	&2.0	& $\nicefrac{1}{5000}$	& 
		2.9508803634	& 3.93620	& 0.4656849766	& 5.69380	& 0.0096895862	& 5.84320	& -0.1118754521	& 0.0033509689	& 5.95620	 &  5083.0
		\\        \hphantom{HDG\textsubscript{$H^{div}$}}\footnotemark[8]        $(L,k)=(3,\hphantom{0}5)$
		&746.0	&8.1	& $\nicefrac{1}{10000}$	&
		2.9508882652	& 3.93620	& 0.4656563266	& 5.69390	& 0.0096885678	& 5.84300	& -0.1118742936	& 0.0033485024	& 5.95630	 &  21829.4
		\\
		\cmidrule(lr){1-14} 
        HDG\textsubscript{$H^{div}$}\footnotemark[5]        $(L,k)=(0,\hphantom{0}5)$
		&19.1	&0.2	& $\nicefrac{1}{1250}$	&
		2.9511386439	& 3.93680	& 0.4756634788	& 5.69280	& 0.0098269810	& 5.84240	& -0.1117569458	& 0.0034432916	& 5.95520	 &  204.9
		\\         \hphantom{HDG\textsubscript{$H^{div}$}}        $(L,k)=(0,\hphantom{0}6)$
		&25.7	&0.3	& $\nicefrac{1}{1644}$	& 
		2.9509517602	& 3.93613	& 0.4697975801	& 5.69161	& 0.0097615210	& 5.84124	& -0.1117081557	& 0.0033587325	& 5.95377	 &  356.4
		\\         \hphantom{HDG\textsubscript{$H^{div}$}}        $(L,k)=(0,\hphantom{0}7)$
		&33.3	&0.4	& $\nicefrac{1}{2071}$	& 
		2.9508984404	& 3.93626	& 0.4668674486	& 5.69339	& 0.0097111637	& 5.84307	& -0.1118381480	& 0.0033566456	& 5.95558	 &  519.4
		\\         \hphantom{HDG\textsubscript{$H^{div}$}}        $(L,k)=(0,\hphantom{0}8)$
		&41.9	&0.5	& $\nicefrac{1}{2530}$	& 
		2.9508992999	& 3.93636	& 0.4661219409	& 5.69368	& 0.0096984067	& 5.84308	& -0.1118513538	& 0.0033486328	& 5.95613	 &  838.7
		\\         \hphantom{HDG\textsubscript{$H^{div}$}}        $(L,k)=(0,\hphantom{0}9)$
		&51.4	&0.6	& $\nicefrac{1}{3019}$  & 
		2.9508944978	& 3.93607	& 0.4658316869	& 5.69361	& 0.0096922436	& 5.84299	& -0.1118662036	& 0.0033490102	& 5.95628	 & 1425.9
		\\         \hphantom{HDG\textsubscript{$H^{div}$}}        $(L,k)=(0,10)$
		&62.0	&0.7	& $\nicefrac{1}{3536}$	& 
		2.9508942137	& 3.93609	& 0.4657339952	& 5.69372	& 0.0096901571	& 5.84304	& -0.1118688505	& 0.0033485887	& 5.95617	 & 2305.8 \\
        \bottomrule        
      \end{tabular}
    }
    \caption{Results for Rot-2D3. The number of degrees of freedom (dof) is given in $K=10^3$. Dofs in brackets are the number of unconstrained dofs after static condensation. The number of non-zero entries (nze) of the condensed (if applicable) system matrix are given in $M=10^6$. The computational run-time is given in seconds $s$.}
	\label{tab:ResultsRot2D3}
	
\end{table}
\footnotetext[7]{Computed using Intel Xeon E5-2640 v4, 2.40GHz using 8 cores.}
\footnotetext[8]{$(L=3,k=5)$: Computed using 2x Intel Xeon E5-2650 v4, 2.20 GHz using 24 cores.} 

\begin{table}
	\centering
	\resizebox{\textwidth}{!}{%
	\begin{tabular}{r ccccc}
		\toprule
		Coefficient & $C_{D,\text{max}}$	&	$t_{D,\text{max}}$	&	$C_{L,\text{max}}$	&	$t_{L,\text{max}}$ & $C_{T,\text{max}}$\\
		\cmidrule(lr){2-6}		
		Value & $[2.95085,2.95092]$ & $[3.9360,3.9372]$ & $[0.4656,0.4659]$ & $[5.6937,5.6951]$ & $[0.0096885,0.009762]$ \\
		\midrule
		Coefficient 	& $t_{T,\text{max}}$	&	$\Delta p(8)$	&	$\omega^\ast_{\text{max}}$&	$t_{\omega,\text{max}}$\\
		\cmidrule(lr){2-6}
		Value & $[5.842,5.845]$ & $[-0.11189,-0.11186]$ & $[0.003347,0.003362]$ & $[5.9558,5.9567]$ \\
		\bottomrule
	\end{tabular}}
	\caption{Reference value ranges for the quantities of interest for the problem Rot-2d3.}
	\label{tab:RefValsRot2D3}
\end{table}


\begin{table}
  \centering
    \resizebox{\textwidth}{!}{%
      \begin{tabular}{crr llllllll r}
        \toprule
        \multicolumn{3}{c}{Discretisation} & \multicolumn{9}{c}{Results}\\
        \cmidrule(lr){1-3}\cmidrule(lr){4-12}
        Method	&	\#dof [$K$]	&	\#nze [$M$]	&	$C_{\Fb_1}$	&	$C_{\Fb_2}$	&	$C_{\Fb_3}$	&	$\Vert C_{\Tb}\Vert_2$	&	$\Delta p$	&	$\angV^\ast_1$&	$\angV^\ast_2$&	$\angV^\ast_3$	& Comp.T. [$s$]	\\
        \midrule
		TH\textsubscript{gd,ho}\footnotemark[9]	
				& 212.9(167.8) 	& 50.9 & 6.02155266	& 0.02502608	& 0.12755555	& 2.09e-12	& 0.13635030	& -2.97751e-06	& 0.03465252	& -0.00689595	& 284\\
		$k=4$	& 652.1(536.3) 	& 158.3 & 6.02151890	& 0.02496250	& 0.12753770	& 6.47e-15	& 0.13681163	& -6.22459e-07	& 0.03464332	& -0.00686038	& 1168 \\
				& 1120.0(923.5) & 272.1 & 6.02151253	& 0.02495675	& 0.12752276	& 1.96e-13	& 0.13662752	& -1.03142e-06	& 0.03464001	& -0.00686132	& 3671 \\
	 			& 1831.0(1468.0) & 440.5 & 6.02150624	& 0.02495834	& 0.12752303	& 3.38e-13	& 0.13661989	& -1.01483e-06	& 0.03463992	& -0.00686150	& 8036 \\	
		\cmidrule(lr){1-12}
		EO\textsubscript{LPS}\footnotemark[3]\textsuperscript{,}\footnotemark[10]
 		&189   & 41    &6.0194099 &0.027620778&0.1413845&5.59e-13&0.1457953&-1.853299-06&0.03793676&-0.00750854&79\\
        &1426  & 336   &6.0209826 &0.025185144&0.1286906&2.52e-09&0.1385932&-1.017284-06&0.03489545&-0.00691099&685\\
        &11061 & 2716  &6.0214724 &0.024973759&0.1276006&3.11e-11&0.1369655&-1.017598-06&0.03465537&-0.00686465&6728\\
        &87100 & 21836 &6.0215013  &0.024959810&0.1275276&6.34e-14&0.1366775&-1.018982-06&0.03464069&-0.00686186&57331\\
        	\cmidrule(lr){2-12}
        	&\multicolumn{2}{l}{extrapolated}              &6.0215031        &0.0249588  &0.1275223  &--&0.1366156&--          &0.03463972&0.006861677\\
                                           &\multicolumn{2}{l}{(order)}           &(4.08)&(3.97)      &(3.92)     &--&(2.50)  &--&(4.03)&(4.05)&\\
        \cmidrule(lr){1-12}
		TH\footnotemark[4] 
		& 37.2 & 3.4 & 5.97793 & 0.0276684 & 0.133363 & 1.77855e-12 & 0.134442  & 0.000106824 & 0.0390861 & -0.00873872 & 26 \\ 
		& 64.4 & 5.8 & 6.00204 & 0.0250739 & 0.126979 & 3.21458e-10 & 0.137206  & 1.47338e-05 & 0.035706 & -0.00759 & 52 \\ 
		& 173.2 & 15.9 & 6.01025 & 0.026336 & 0.128098 & 1.65002e-10 & 0.135844  & -1.99197e-05 & 0.0353658 & -0.00765735 & 195 \\ 
		& 285.1 & 25.9 & 6.01655 & 0.0253422 & 0.128058 & 7.69239e-10 & 0.13658  & 1.35604e-05 & 0.0353014 & -0.00720906 & 435 \\ 
		& 990.2 & 93.1 & 6.01912 & 0.0249711 & 0.127594 & 1.85239e-10 & 0.136474  & 1.2935e-06 & 0.0348498 & -0.00691854 & 3587 \\   
        \cmidrule(lr){1-12}
		HDG\textsubscript{$H^{div}$}\footnotemark[11]
		$k=1$ & 133.4	(116.0) & 7.7	& 5.26348 & 0.0557553 & 0.115768 & 7.77e-13 & 0.137416 & 0.0092711    & 0.0341869 & -0.0268611  & 36  \\
		$k=2$ & 320.7	(232.0) & 30.7	& 6.23837 & 0.0411259 & 0.143691 & 4.21e-12 & 0.139044 & 0.000612444  & 0.0384272 & -0.0125745  & 118 \\
		$k=3$ & 624.3	(386.7) & 85.4	& 6.00429 & 0.0283705 & 0.125754 & 4.61e-13 & 0.132603 & -0.000164957 & 0.0345766 & -0.00779181 & 545 \\
		$k=4$ & 1071.0	(580.1) & 192.1	& 6.01982 & 0.0251    & 0.126563 & 1.39e-12 & 0.13611  & 5.29841e-06  & 0.0343653 & -0.00686743 & 752 \\
		$k=5$ & 1688.0	(812.1) & 376.5	& 6.02186 & 0.0248633 & 0.127461 & 8.03e-12 & 0.13656  & -1.42517e-05 & 0.0346083 & -0.00684754 & 2319\\
		$k=6$ & 2501.9	(1082.8) & 669.3 & 6.02153 & 0.0249391 & 0.127536 & 6.12e-12 & 0.136585	& -1.25483e-06 &0.0346432 &	-0.00685716 & 4122 \\
        \bottomrule        
      \end{tabular}
    }
    \caption{Results for Rot-3d1. The number of degrees of freedom (dof) is given in $K=10^3$. Dofs in brackets are the number of unconstrained dofs after static condensation. The number of non-zero entries (nze) of the condensed (if applicable) system matrix are given in $M=10^6$. The computational run-time is given in seconds $s$.}
    \label{tab:ResultsRot3D1}
\end{table}
\footnotetext[9]{Computed using 2x Intel Xeon E5-2650 v4, 2.20GHz using 24 cores.}
\footnotetext[10]{Computed using 2x Intel Xeon E5-2640 v4, 2.40GHz using 20 cores.} 
\footnotetext[11]{$1 \leq k \leq 3$: Computed using Intel Xeon E5-2650 v4, 2.20 GHz using 12 cores, \\
  $k = 4$: Computed using 2x Intel Xeon E5-2650 v4, 2.20 GHz using 24 cores,\\
  $k = 5$: Computed using 4x Intel Xeon E5-4620 v3, 2.00 GHz using 18 cores,\\
  $k = 6$: Computed using 4x Intel Xeon E5-4620 v3, 2.00 GHz using 24 cores.}

\begin{table}
	\centering
	\begin{tabular}{r cccc}
		\toprule
		Coefficient	& $C_{\Fb_1}$	&	$C_{\Fb_2}$	&	$C_{\Fb_3}$	&	$\Delta p$	\\
		\cmidrule(lr){2-5}
		Value 	& $[6.019,6.0216]$	 & $[0.02493,0.02498]$ & $[0.12752,0.1276]$ & $[0.1364,0.1367]$ \\
		\midrule
		Coefficient&	$\angV^\ast_1$&	$\angV^\ast_2$&	$\angV^\ast_3$ \\
		\cmidrule(lr){2-5}
		Value	& $[-1.3\cdot10^{-6},1.3\cdot10^{-6}]$ & $[0.034639,0.03485]$ & $[-0.00692,-0.00685]$\\
		\bottomrule
	\end{tabular}
	\caption{Reference Values for the problem Rot-3d1.}
	\label{tab:RefValsRot3D1}
\end{table}

%% file: plot2d2.tex
\pgfplotsset{
    every non boxed x axis/.style={} 
}
\definecolor{color0}{rgb}{0.12156862745098,0.466666666666667,0.705882352941177}

\begin{tikzpicture}
	\begin{groupplot}[
		group style = {
			group size = 1 by 5,
			xticklabels at=edge bottom,
			vertical sep= 0pt},
			width = \textwidth,
			height= 2\textwidth,
		]
		\nextgroupplot[
			ylabel = {~~Drag},
			height = 4cm,
			axis x line*=top,
			xtick={28.044,28.376,28.707},
			axis y discontinuity=parallel,
			ytick={3.23, 3.20, 3.17},
			ymin = 3.15, ymax= 3.24,
		]			]
		\addplot +[mark=none, color0] table [x=Time, y=Drag, col sep=tab] {Rot2D2-Functionals-sample.txt};
		\nextgroupplot[
			ylabel = {~~~~Lift},
			ylabel shift={5pt},
			height = 4cm,
			axis x line=none,
			axis y discontinuity=parallel,
			ymin=-1.5, ymax = 1.1,
			]
		\addplot +[mark=none, color0] table [x=Time, y=Lift, col sep=tab] {Rot2D2-Functionals-sample.txt};
		\nextgroupplot[
			ylabel = {~~~~Pressure difference},
			ylabel shift={1pt},
			height = 4cm,
			axis x line=none,
			axis y discontinuity=parallel,
			ymin  = 2.375, ymax = 2.51 
			]
		\addplot +[mark=none, color0] table [x=Time, y=pDiff, col sep=tab] {Rot2D2-Functionals-sample.txt};
		\nextgroupplot[
			ylabel = {~~~Torque},
			ylabel shift={3pt},
			height = 4cm,
			axis x line=none,
			axis y discontinuity=parallel,
			ytick={0.02, 0.01,0,-0.01},
			ymin=-0.025,
			every y tick scale label/.style={at={(rel axis cs:0.25,1.12)},anchor=north east},
			]
		\addplot +[mark=none, color0] table [x=Time, y=Torque, col sep=tab] {Rot2D2-Functionals-sample.txt};	
		\nextgroupplot[
			ylabel = {~~Angular velocity},
			ylabel shift={9pt},
			height = 4cm,
			axis x line*=bottom,
			xlabel= {Time},
			xtick={28.044,28.376,28.707},
			xticklabels={$t_0$, $t_0+\nicefrac{1}{f}$, $t_0+\nicefrac{2}{f}$},
			every y tick scale label/.style={at={(rel axis cs:0.25,1.12)},anchor=north east},
			]
		\addplot +[mark=none, color0] table [x=Time, y=AngVel, col sep=tab] {Rot2D2-Functionals-sample.txt};		
	\end{groupplot}
\end{tikzpicture}

%% file: plot2d3.tex
\pgfplotsset{
    every non boxed x axis/.style={} 
}
\definecolor{color0}{rgb}{0.12156862745098,0.466666666666667,0.705882352941177}

\begin{tikzpicture}
	\begin{groupplot}[
		group style = {
			group size = 1 by 5,
			xticklabels at=edge bottom,
			vertical sep= 0pt},
			width = \textwidth,
			height= 2\textwidth,
		]
		\nextgroupplot[
			ylabel = {~~Drag},
			height = 4cm,
			ylabel shift={6pt},
			axis x line*=top,
			axis y discontinuity=parallel,
			ymin=-0.8
			]
		\addplot +[mark=none, color0] table [x=Time, y=Drag, col sep=tab] {Rot2D3-Functionals-sample.txt};
		\nextgroupplot[
			ylabel = {~~~~Lift},
			ylabel shift={-5pt},
			height = 4cm,
			axis x line=none,
			axis y discontinuity=parallel,
			ymin=-0.74, ymax=0.52
			]
		\addplot +[mark=none, color0] table [x=Time, y=Lift, col sep=tab] {Rot2D3-Functionals-sample.txt};
		\nextgroupplot[
			ylabel = {~~~~Pressure difference},
			ylabel shift={8pt},
			height = 4cm,
			axis x line=none,
			axis y discontinuity=parallel,
			ymin=-0.65
			]
		\addplot +[mark=none, color0] table [x=Time, y=pDiff, col sep=tab] {Rot2D3-Functionals-sample.txt};
		\nextgroupplot[
			ylabel = {~~~Torque},
			height = 4cm,
			axis x line=none,
			axis y discontinuity=parallel,
			every y tick scale label/.style={at={(rel axis cs:0.25,1.12)},anchor=north east},
			ymin= -0.015,
			]
		\addplot +[mark=none, color0] table [x=Time, y=Torque, col sep=tab] {Rot2D3-Functionals-sample.txt};	
		\nextgroupplot[
			ylabel = {~~Angular velocity},
			ylabel shift={7pt},
			height = 4cm,
			axis x line*=bottom,
			xlabel= {Time},
			every y tick scale label/.style={at={(rel axis cs:0.25,1.12)},anchor=north east},
			]
		\addplot +[mark=none, color0] table [x=Time, y=AngVel, col sep=tab] {Rot2D3-Functionals-sample.txt};		
	\end{groupplot}
\end{tikzpicture}

%% file: back_matter.tex
\section{Code availability}

Full data sets for the results presented in Section~\ref{sec:Results} as well as code used to compute these benchmarks can be found at
\vskip .5\baselineskip
\centerline{\href{https://doi.org/10.5281/zenodo.3253455}{DOI: \texttt{10.5281/zenodo.3253455}}.}
\vskip .5\baselineskip
\noindent Here we provide both the scripts used to compute the respective benchmarks and instructions and links to the relevant external software used.

\section*{Acknowledgements}

HvW, TR and PM acknowledge support by the Deutsche Forschungsgemeinschaft (DFG, German Research Foundation) - 314838170, GRK 2297 MathCoRe. TR and PM further acknowledge supported by the Federal Ministry of Education and Research of Germany (project number 05M16NMA).